\documentclass[aps,prf,onecolumn,groupedaddress,superscriptaddress,notitlepage,longbibliography]{revtex4-1}
\usepackage[utf8]{inputenc}
\usepackage{amsmath, graphicx}
\usepackage{graphicx}
\usepackage{dcolumn}
\usepackage{bm}
\usepackage{natbib}
\usepackage{epstopdf}
\usepackage{natbib}
\usepackage{amsmath}
\usepackage[abs]{overpic}
\usepackage{subfigure}
\usepackage{color}
\begin{document}
\author{Christian Pedersen}
\affiliation{Mechanics Division, Department of Mathematics, University of Oslo, 0316 Oslo, Norway.}
\author{John Niven}
\affiliation{Department of Physics and Astronomy, McMaster University, 1280 Main Street West, Hamilton, Ontario, L8S 4M1, Canada.}
\author{Thomas Salez}
\affiliation{Univ. Bordeaux, CNRS, LOMA, UMR 5798, F-33405 Talence, France.}
\affiliation{Global Station for Soft Matter, Global Institution for Collaborative Research and Education, Hokkaido University, Sapporo, Hokkaido 060-0808, Japan.}
\author{Kari~Dalnoki-Veress}
\affiliation{Department of Physics and Astronomy, McMaster University, 1280 Main Street West, Hamilton, Ontario, L8S 4M1, Canada.}
\author{Andreas Carlson}
\email{acarlson@math.uio.no}
\affiliation{Mechanics Division, Department of Mathematics, University of Oslo, 0316 Oslo, Norway.}
\title{Asymptotic regimes in elastohydrodynamic and stochastic leveling on a viscous film}
\date{\today}
\begin{abstract}
An elastic sheet that deforms near a solid substrate in a viscous fluid is a situation relevant to various dynamical processes in biology, geophysics and engineering. Here, we study the relaxation dynamics of an elastic plate resting on a thin viscous film that is supported by a solid substrate. By combining scaling analysis, numerical simulations and experiments, we identify asymptotic regimes for the elastohydrodynamic leveling of a surface perturbation of the form of a bump, when the flow is driven by either the elastic bending of the plate or thermal fluctuations. In both cases, two distinct regimes are identified when the bump height is either much larger or much smaller than the thickness of the pre-wetted viscous film. Our analysis reveals a distinct crossover between the similarity exponents with the ratio of the perturbation height to the film height.
\end{abstract}
\maketitle

\section{INTRODUCTION}
The motion of an elastic sheet supported by a thin layer of viscous fluid is a phenomenon that manifests itself in processes spanning wide ranges of time and length scales, from \textit{e.g.} magmatic intrusion in the Earth's crust~\cite{balmforth2005instability,michaut2011dynamics}, to fracturing and crack formation in glaciers~\cite{ball2018static}, to pumping in the digestive and arterial systems~\cite{bilston2003arterial, takagi2011peristaltic, borcia2018liquid}, or the construction of 2D crystals for electronic engineering~\cite{sanchez2018mechanics}. Elastohydrodynamic flows have been studied in model geometries in order to understand their generic features and the inherent coupling between the driving force from the elastic deformations of the material and the viscous friction force resisting motion~\cite{hosoi2004peeling,Huang2002,al2013two,hewitt2015elastic,tulchinsky2016transient,elbaz2016axial,arutkin2017elastohydrodynamic,juel2018instabilities, rivetti2017elastocapillary}.

The investigation of an initially flat elastic membrane that is subsequently subjected to an applied deformation has helped disclose how system size, magnitude and direction of elastic deformations and spatial confinement affect the membrane dynamics. Whilst stretching~\cite{bernal2011elastic, box2018impact} or compressing~\cite{Vandeparre2010, Kodio2017} of the membrane produces wrinkling patterns~\cite{cerda2003geometry}, it has been shown that slowly deforming the membrane, by means of injecting additional fluid into the supporting layer, leads to a dynamics where the fluid pressure is solely balanced by elastic bending forces \cite{lister2013viscous, berhanu2019uplift}. As the out-of-plane deflection increases, a change in the physical mechanism that dictates the dynamics occurs as elastic stretching becomes the dominant driving mechanism until the system reaches a critical size for which gravity starts to dominate the dynamics. For small systems, an analog to such a predicted transition is observed for a thin perturbed elastic plate resting on a nanoscopic fluid layer, where the restoring elastic bending force is opposed by van der Waals forces leading to an elastohydrodynamic touchdown~\cite{carlson2016similarity} similar to capillary film dewetting~\cite{zhang1999similarity}.
To describe the dynamics theoretically, one can solve the full Navier-Stokes equation in the fluid phase using dynamic boundary conditions at the elastic interface given by the solution of the F\"oppl-von-K\'arm\'an equation~\cite{pihler2015displacement}, using \textit{e.g.} the immersed-boundary method~\cite{heil2011fluid}. Viscous flow in thin films can be described by the lubrication theory~\cite{batchelor2000introduction} that has been widely used to study different elastohydrodynamic flow phenomena \cite{hosoi2004peeling,lister2013viscous, berhanu2019uplift,pihler2015displacement,carlson2016similarity}. However, not much is known about how elastohydrodynamic flows are affected by the ratio between the geometric parameters that characterize the system as it undergoes large changes while the driving force remain the same.

For instance, when an elastic sheet deforms onto a wall pre-wetted by a thin viscous film, the dynamics of the advancing front are dictated by the local curvature of the interface~\cite{lister2013viscous, rivetti2017elastocapillary}. This elastohydrodynamic relaxation is reminiscent of the capillary spreading of a viscous drop onto a solid substrate~\cite{tanner1979spreading,de1985wetting, cormier2012beyond, bergemann2018viscous}. Similar to capillary flows, elastohydrodynamic relaxation processes are not only limited to very thin pre-wetted films. In fact, an elastic sheet with zero spontaneous curvature but with an initial shape of a bump (Fig.~\ref{fig:illustration}) with a height much larger than the pre-wetted viscous film will relax towards a flat equilibrium state. Inevitably, the system must then crossover from a situation where the bump height is larger than the pre-wetted film height, to a situation where instead the pre-wetted film becomes thicker than the bump. Here we investigate how the elastohydrodynamic leveling changes with the ratio between the bump height and the pre-wetted film thickness. In particular, are there different asymptotic regimes, and how do the system transition from one to another? At the nanoscale, thermal fluctuations are expected to contribute and may even dominate the dynamics~\cite{davidovitch2005spreading,Demery2011,nesic2015fully,carlson2018fluctuation,marbach2018transport}, which we quantify in the leveling dynamics. To answer these questions, we combine numerical solutions of a mathematical model based on the lubrication theory~\cite{batchelor2000introduction} with scaling analysis and experiments.

\section{MATHEMATICAL MODEL AND NUMERICAL PROCEDURE}
\begin{figure}
\center
\includegraphics[width=0.99\linewidth]{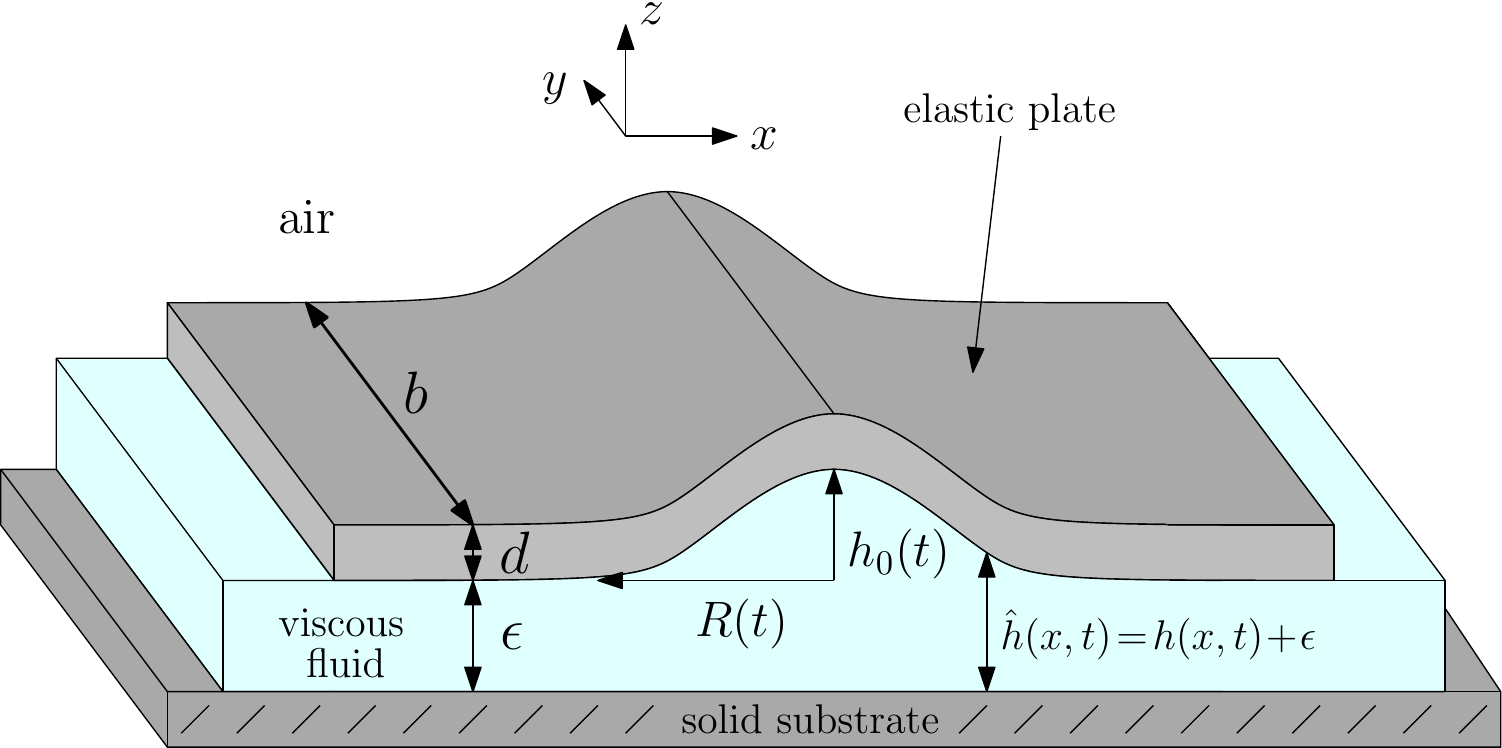}
\caption{Schematic of the system we are studying, where an elastic plate is supported by a viscous film on a solid substrate, surrounded by ambient air. The elastic plate has a thickness $d$ and a width $b$. Initially, the overall profile presents a localized bump, whose profile is invariant in the $y$-direction, \textit{i.e.}, quasi two-dimensional. Far away from the perturbation, the viscous film has a constant thickness $\epsilon$. In the bump region, the height profiles $\hat{h}(x,t)$ and $h(x,t)=\hat{h}(x,t)-\epsilon$ of the viscous film and the bump, respectively, vary with the horizontal position $x$ and time $t$, and remain symmetric about $x=0$. At $x=0$, we define the characteristic height $h(x=0,t)=h_0(t)$ of the bump and its typical radius $R(t)$, with initial values given as $h_0(t=0)=h_{\textrm{i}}$ and $R(t=0)=R_{\textrm{i}}$.}
\label{fig:illustration}
\end{figure}
We consider the system depicted in Fig.~\ref{fig:illustration}, where we focus on a system where any influence of gravity can be neglected, \textit{i.e.}, the bump height is smaller than the elasto-gravity length \cite{hosoi2004peeling}. Only situations where the bump height $h(x,t)$ is small compared with its horizontal extent and where the film slopes are small, \textit{i.e.}, $\partial \hat{h}(x,t)/\partial x\ll 1$, are considered. We describe the viscous flow between the plate and the solid substrate using lubrication theory~\cite{batchelor2000introduction}. When the initial deflection $h_{\textrm{i}}$ of the elastic plate is small compared to its thickness $d$, we can neglect stretching and the pressure reduces to $p(x,t) = B\partial ^4 \hat{h}(x,t)/\partial x^4$, where $B=Ed^3/[12(1-\nu^2)]$ is the bending rigidity of the plate, $E$ is the Young modulus and $\nu$ is the Poisson's ratio~\cite{landau1959course}. In addition, the system is a spatially unconfined elastic sheet with the two lateral boundaries being free to move relative to the underlying fluid. Thus, the in-plane compression is suppressed, and bending stresses dominate the relaxation process regardless of the ratio $d/h_{\textrm{i}}$. By assuming incompressible flow and imposing no-slip conditions at the two solid substrates, and considering a one-dimensional geometry as there are no variations along the $y$-direction, one obtains the governing equation for the evolution of the height profile (see \textit{e.g.} Ref.~\cite{hosoi2004peeling})
\begin{equation}
\frac{\partial \hat{h}(x,t)}{\partial t} = \frac{\partial}{\partial x}\left[\frac{B}{12\mu}\hat{h}^3(x,t)\frac{\partial^5 \hat{h}(x,t)}{\partial x^5} + \Gamma \hat{h}^{3/2}(x,t)\eta(x,t)\right]\ ,
\label{eq:evolution_equation}
\end{equation}
where $\mu$ is the fluid's dynamic viscosity. At small scales, thermal fluctuations can also influence the dynamics, which is described by the last term of Eq.~\eqref{eq:evolution_equation}. This term mimics the stress generated by thermal fluctuations, originates from an additional symmetric random stress term in the Navier-Stokes equations and is obtained by an integration in the $z$-direction (for details see~\cite{grun2006thin, duran2019instability, nesic2015dynamics, carlson2018fluctuation}). The noise term $\eta(x,t)$, is multiplied by a prefactor $\Gamma=\sqrt{k_{\textrm{B}}T_{\textrm{A}}/(6\mu b)}$ where $k_{\textrm{B}}$ is the
Boltzmann constant, $T_{\textrm{A}}$ is the ambient temperature, $b$ is the width of the plate along the $y$-direction, and $\eta(x,t)$ is modelled as a spatiotemporal Gaussian white noise such that $\langle\eta(x,t)\rangle=0$ and $\langle\eta(x,t)\eta(x',t')\rangle=\delta(x-x')\delta(t-t')$, where the $\langle ~\rangle$ symbols indicate average quantities. We non-dimensionalize Eq.~\eqref{eq:evolution_equation} by: $X=x/R_{\textrm{i}}$, $\hat{H}(X,T)=\hat{h}(x,t)/h_{\textrm{i}}$, $T= t Bh_{\textrm{i}}^{\,3} /(12\mu R_{\textrm{i}}^{\,6})$, and $\Theta(X,T)=\eta(x,t)\sqrt{12\mu R_{\textrm{i}}^{\,7}/(Bh_{\textrm{i}}^{\,3})}$. When $\Gamma=0$, this non-dimensionalization procedure gives us a parameter-free partial differential equation for $\hat{H}(X,T)$. When $\Gamma >  0$, the non-dimensional
number $N=\sqrt{2k_{\textrm{B}}T_{\textrm{A}}R_{\textrm{i}}^{\,3}/(Bh_{\textrm{i}}^{\,2} b)}$ appears as a pre-factor in front of the stochastic term, and $N^2$ measures the ratio between thermal and bending energies. For the macroscopic system provided in our experiment and described in detail below, \textit{i.e.}, $T_A = 300$ K, $h_{\textrm{i}}=2.5\; \mu\textrm{m}$, $R_{\textrm{i}}=20\; \mu\textrm{m}$, $\mu=10^4\; \textrm{Pa}\;\textrm{s}$ and $B=1.3\cdot 10^{-12}\;\textrm{Nm}$ we get the noise prefactor $\Gamma = 2.5\cdot 10^{-13}\; \textrm{m}\textrm{s}^{-1/2}$ and the energy ratio $N=1.75\cdot 10^{-6}$ which is well within the elastic bending dominated regime. However, a transition from a dominant elastohydrodynamic leveling to a dominant stochastic leveling would occur for a system with temperature $T_{\textrm{A}}=300\;$K, membrane perturbation height $h_{\textrm{i}}=10\;$nm and radius $R_{\textrm{i}}=5\;\mu$m for a bending modulus $B$ in the range of $10-100\;k_{\textrm{B}}T_{\textrm{A}}$ where $k_{\textrm{B}}T_{\textrm{A}}=4\times 10^{-21}\;\textrm{Nm}$ which corresponds to $N$ in the range of $0-8$ \cite{carlson2018fluctuation}.

We solve the dimensionless version of Eq.~\eqref{eq:evolution_equation} numerically by using a finite element method, and we split it into three coupled equations for: the bump profile $H(X,T)=\hat{H}(X,T)-\epsilon/h_{\textrm{i}}$, the linearized curvature $\partial^2 H(X,T)/\partial X^2 $, and the bending pressure $\partial^4 H(X,T)/\partial X^4 $. These fields are discretized with linear elements and solved by using Newton's method from the FEniCS library~\cite{logg2012automated}. For the deterministic case $N=0$, an adaptive time stepping routine has been used with an upper time step limit of $\Delta T=0.001$ and a discretization in space $\Delta X \in [0.001; 0.01]$. For the stochastic case $N> 0$, we have used a constant time step $\Delta T=0.001$, together with a discretization in space $\Delta X = 0.0025$. At $T=0$ we impose the initial condition: $H(X,T=0) = 1 - \textrm{tanh}(50X^2)$. We further impose the following boundary conditions at the boundary $\partial\Omega$ of the numerical domain: $H\big(X\in\partial\Omega, T)=H\big(X\in\partial\Omega, 0)$, $\partial ^2 H(X\in\partial\Omega, T)/\partial X^2=0$, and $\partial ^4 H(X\in\partial\Omega, T)/\partial X^4=0$.
 The noise $\Theta(X,T)$ is introduced independently at each discrete position and time step using the ``random'' class with the ``randn'' Gaussian subclass from the Numpy library~\cite{numpy}, with zero mean and a variance $1/(\Delta X \Delta T)$. We avoid negative values of $\hat{H}(X,T)$ (that might occur in the stochastic case due to the fluctuations), by imposing that when $\hat{H}(X,T) < 10^{-6}$, it is put back to $10^{-6}$ as in~\cite{davidovitch2005spreading, nesic2015fully}.
 To verify the predictions of Eq.(1), we construct an experimental setup which is described in the following section.

\section{EXPERIMENTAL PROCEDURE}

The experimental setup is composed of a fiber of polystyrene (PS) with a glass-transition temperature $T_{\text{g, PS}} \approx 100^{ \circ}$C deposited on a film of the same polymer supported on a silicon (Si) substrate. These samples are capped by a thin sheet of polysulfone (PSU) with $T_{\text{g, PSU}} \approx 180\,^{\circ}$C. Sample preparation is carried out as follows: PS fibers (with number-averaged molecular weight $M_{\text{n}} = 15.8$~kg/mol, and polydispersity index PDI = 1.05, Polymer Source Inc., Canada) are pulled from the melt at 175$^{\circ}$C using a glass rod. Thin PS films are spin cast from a toluene solution onto 10$\times$10 mm$^2$ Si substrates, leading to a thickness of 25 to 380~nm, measured using ellipsometry (Accurion, EP3). The films are annealed at 110$^{\circ}$C for at least 12 hours in vacuum to remove residual solvent and relax residual stresses. The PS fibers are then transferred onto the PS films and the ensemble is heated briefly above $T_{\text{g, PS}}$. The heating allows the PS to flow, thereby resulting in a bump. Thin PSU films ($M_{\text{n}} \approx$ 22 kg/mol, Sigma-Aldrich) are prepared by spin casting from a cyclohexanone solution onto freshly cleaved mica substrates (Ted Pella, USA). The PSU films have a thickness of $\approx$ 160 nm, measured using ellipsometry, and are annealed in vacuum at 200$^{\circ}$C for at least 12 hours. The PSU films are floated onto water and then transferred onto a supporting apparatus (described previously~\cite{schulman2018}), held only by the film edges. These freestanding films can be relaxed to an unstrained state ensuring no in-plane tension. The PSU films are finally transferred onto the PS sample. The part of PSU film at the edges of the Si wafer was then removed using a scalpel blade prior to annealing. This was done to ensure slippage at the boundary between the PSU film and liquid PS layer, thus rendering the relaxation bending dominated as discussed above.

\begin{figure}
\includegraphics[trim= 0cm 0cm 0cm 0cm,width=0.8\linewidth]{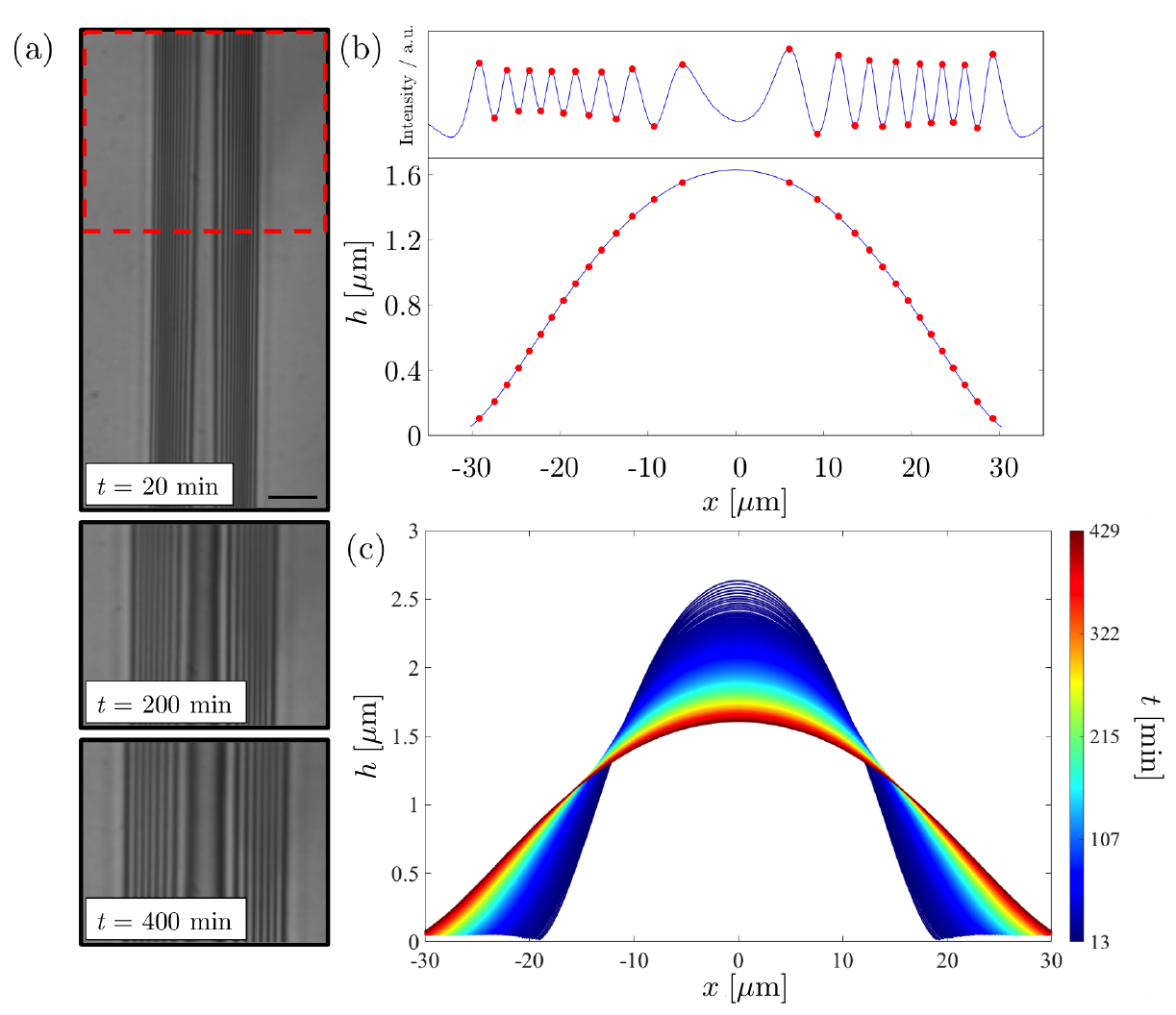}
\caption{(a) Typical optical microscopy images showing the temporal evolution of the interference fringes due to the liquid bump capped by the elastic plate (20 $\mu$m scale bar). The image at 20 minutes is uncropped, showing the invariance in the $y$-direction, while the later images are cropped at the red box. (b) Intensity profile (averaged along the $y$-direction) of the bump at a given time $t$, and corresponding reconstructed bump profile at 400 minutes. (c) Temporal evolution of the bump profile.}
\label{fig:exp_prof}
\end{figure}

After preparation, the samples were annealed on a hotstage (Linkam, UK) at $130^{\circ}$C, which is above $T_{\text{g, PS}}$ but below $T_{\text{g, PSU}}$. Hence, the PS becomes a viscous liquid while the capping PSU film remains an elastic solid, thus realizing the system illustrated in Fig.~\ref{fig:illustration}. The height profile is imaged during annealing using optical microscopy with a red laser line filter ($\lambda$ = 632.8 nm, Newport, USA), which creates interference fringes in the region of the bump, as shown in Fig.~\ref{fig:exp_prof}(a), due to the light that is reflected from the Si substrate. It is clear from these fringes that the initial fiber and resulting flow are one-dimensional over length scales that are many times the width of the perturbation itself. Each interference fringe corresponds to a change in height of $\lambda/(4n)$, where $n \approx 1.57$ is the average index of refraction of the two polymers that make up the sample ($n_{\text{PS}}$ = 1.53 and $n_{\text{PSU}}$ =1.61). This allows the bump profile $h(x,t)$ to be reconstructed by fitting a polynomial to the fringe data, as shown in Fig.~\ref{fig:exp_prof}(b). Such profiles can then be used to track the leveling dynamics, and to extract in particular the evolution of the height $h_0(t)$ of the bump with time for various initial geometries.

\section{RESULTS}
\subsection{Elastohydrodynamic leveling}
\begin{figure}
\subfigure[ ]{\includegraphics[width=.49\textwidth]{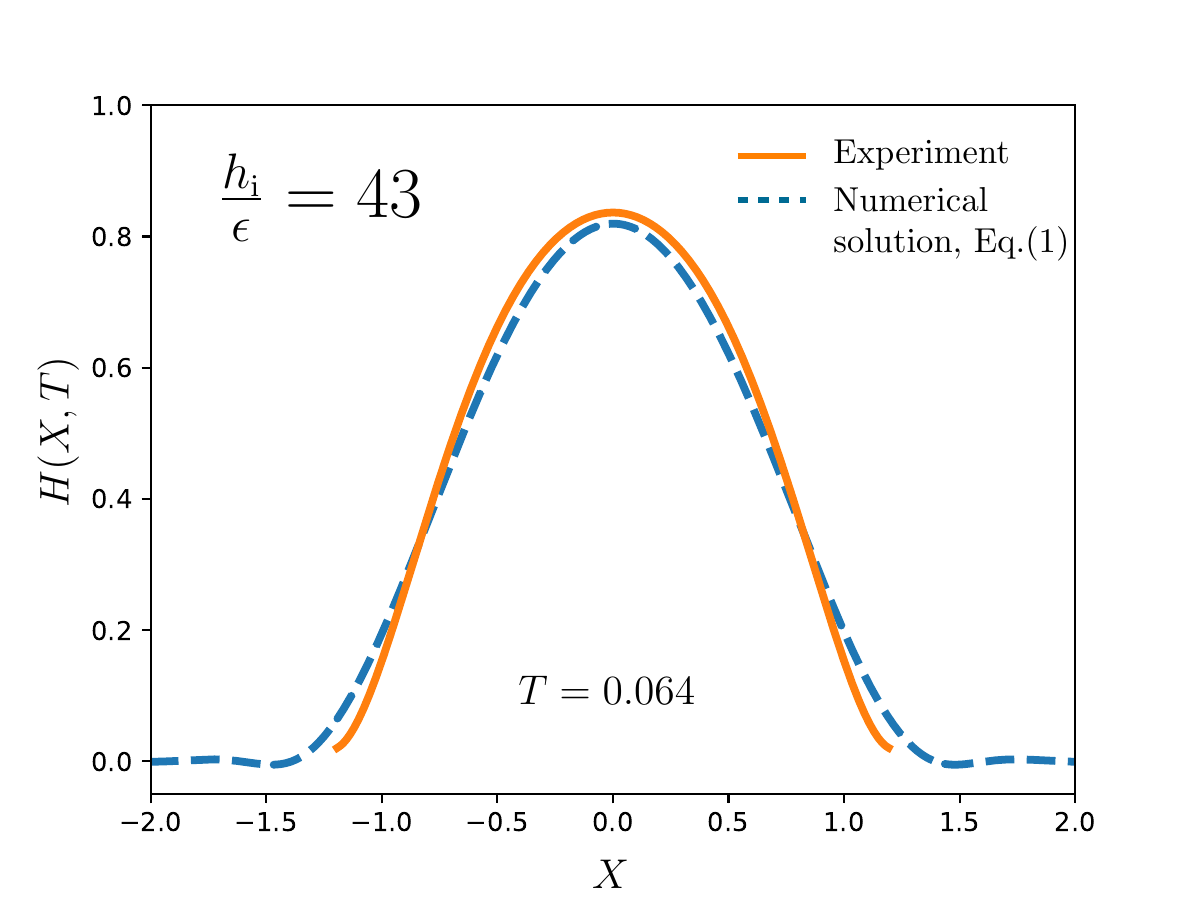}}
\subfigure[ ]{\includegraphics[width=.49\textwidth]{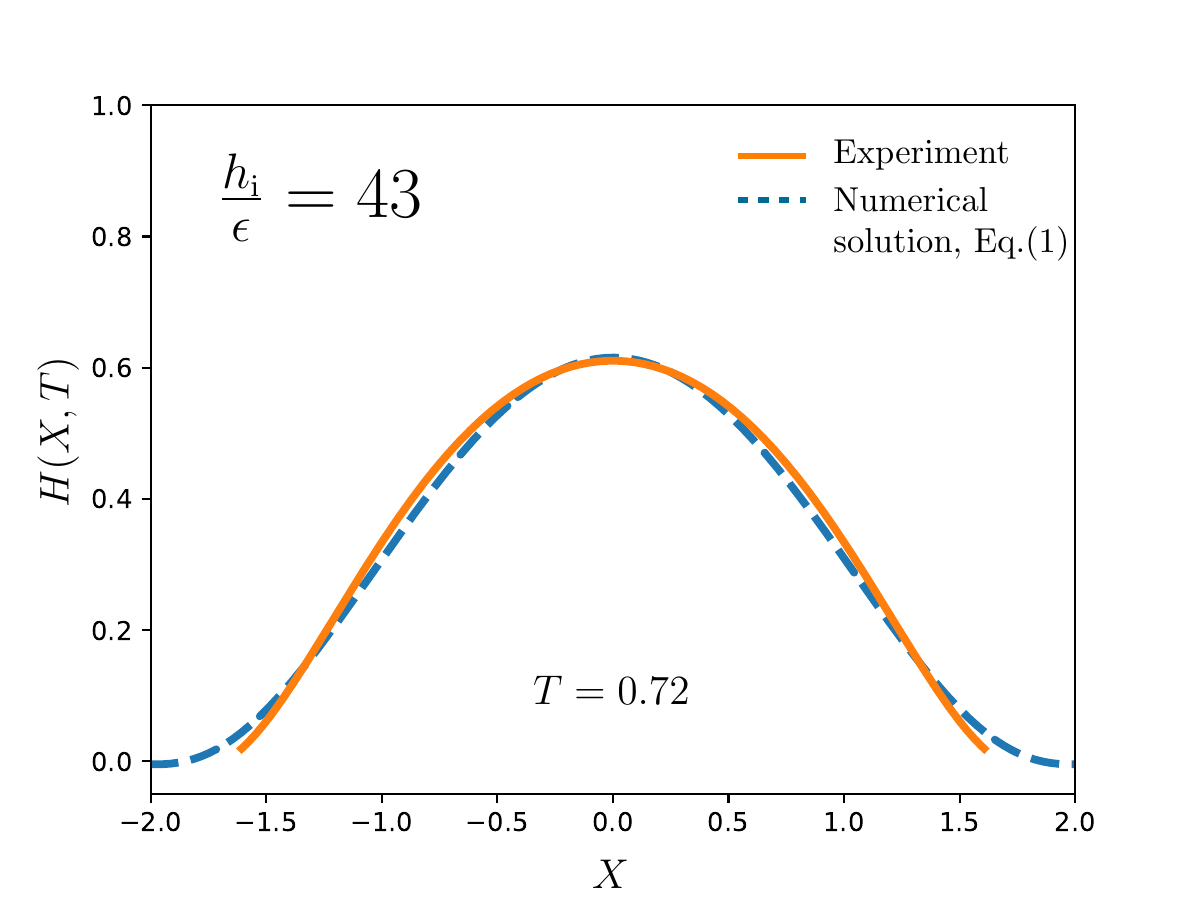}}
\subfigure[]{\includegraphics[width=0.49\linewidth]{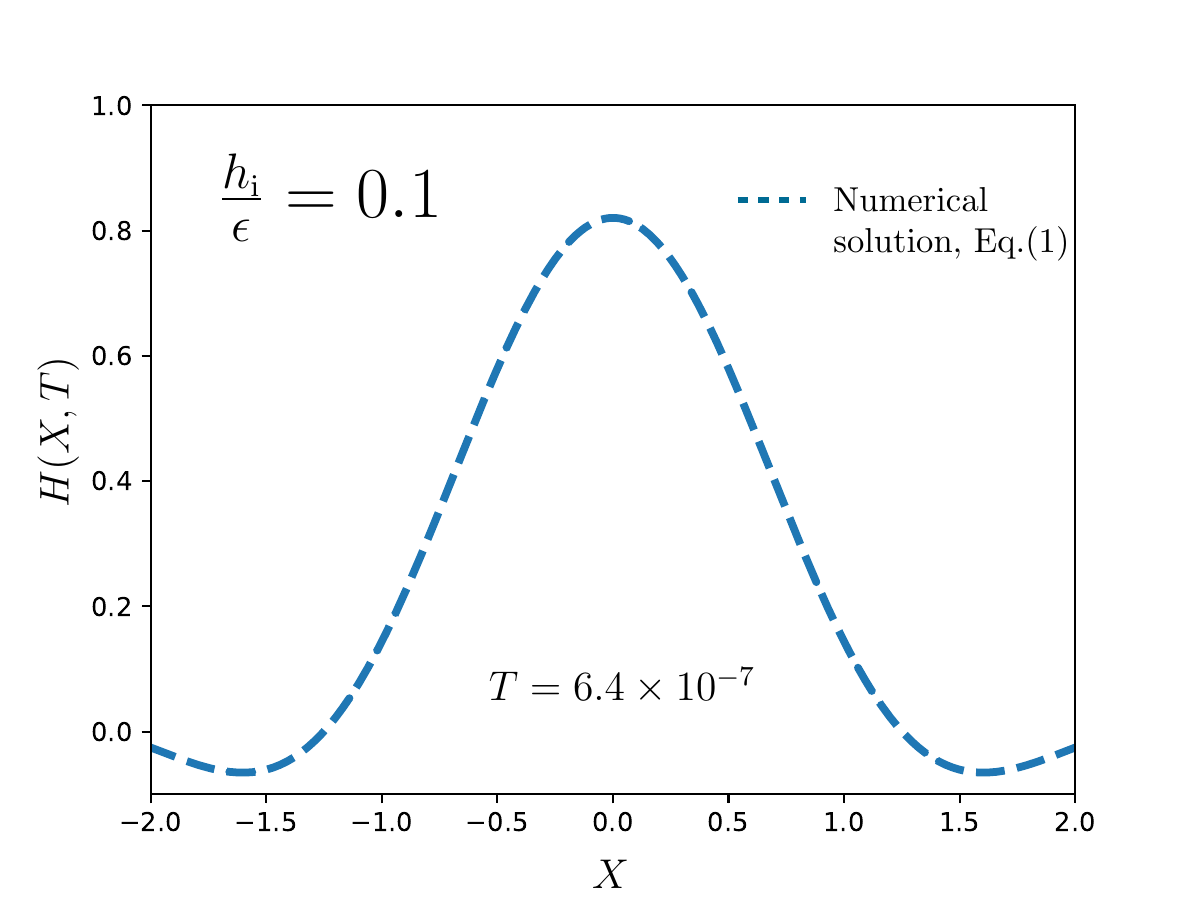}}
\subfigure[]{\includegraphics[width=0.49\linewidth]{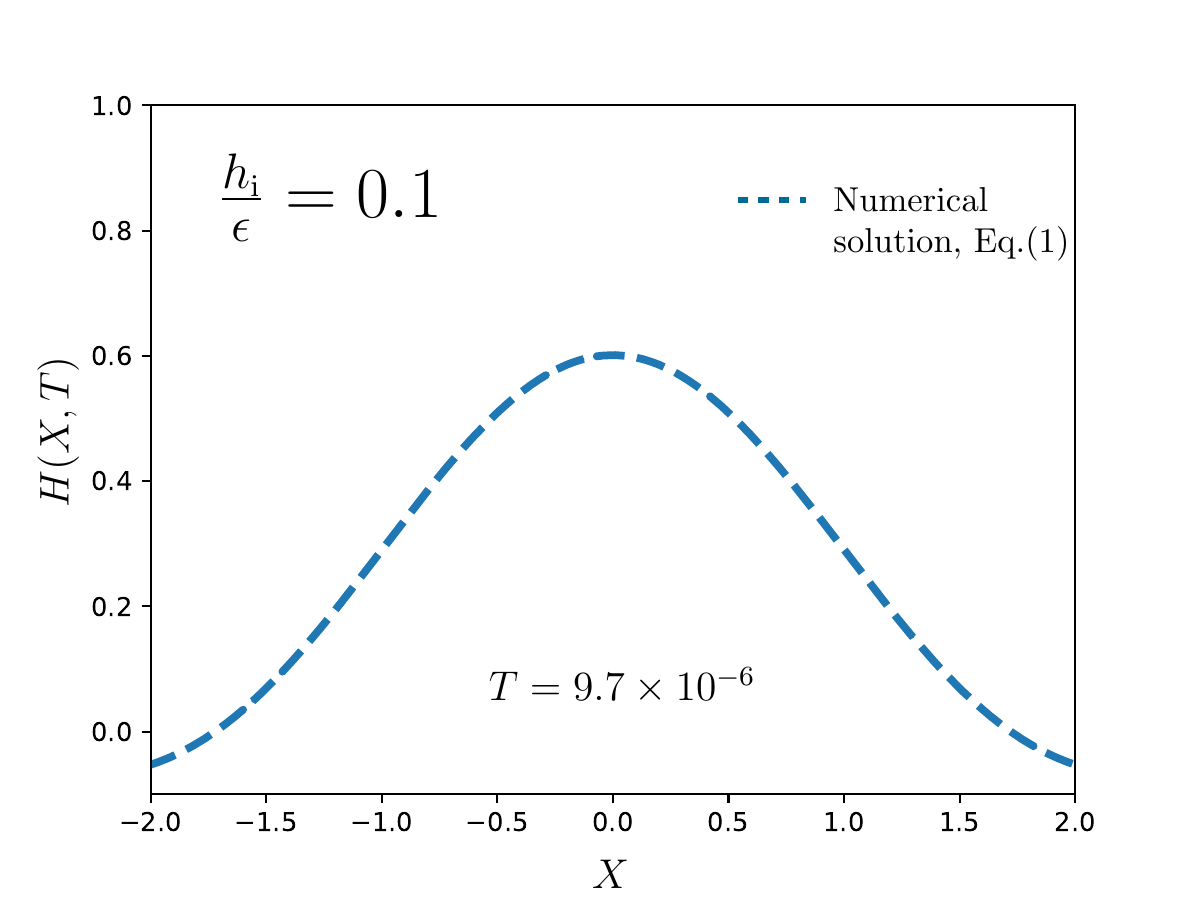}}
\caption{(a-b) Bump height profiles for an initial aspect ratio $h_{\textrm{i}}/\epsilon=43$, from a numerical solution and an experiment at two dimensionless times $T$ as indicated, which correspond respectively to $t=37$ min and $t=416$ min. The geometrical parameters are $h_{\textrm{i}}$ = 2.9 $\mu$m, $R_{\textrm{i}}$ = 16.5 $\mu$m, and $\epsilon$ = 67 nm. In order to account for the experimental uncertainties in the geometrical parameters, the experimental time $t$ is divided by a free fitting factor $\alpha=0.13$. Note that, specifically for these figures, the initial condition for the numerical solution was fixed by a curve fitting of the actual experimental profile at $t=13$min. (c-d) Bump height profiles for an initial aspect ratio $h_{\textrm{i}}/\epsilon =0.1$, from a numerical solution at two dimensionless times $T$, as indicated, chosen so that the central heights $H(X=0,T)$ match the ones in the top row.}
\label{fig:num_vs_exp}
\end{figure}

\begin{figure}
\includegraphics[trim= 3 3 27 20, clip, width=1\linewidth]{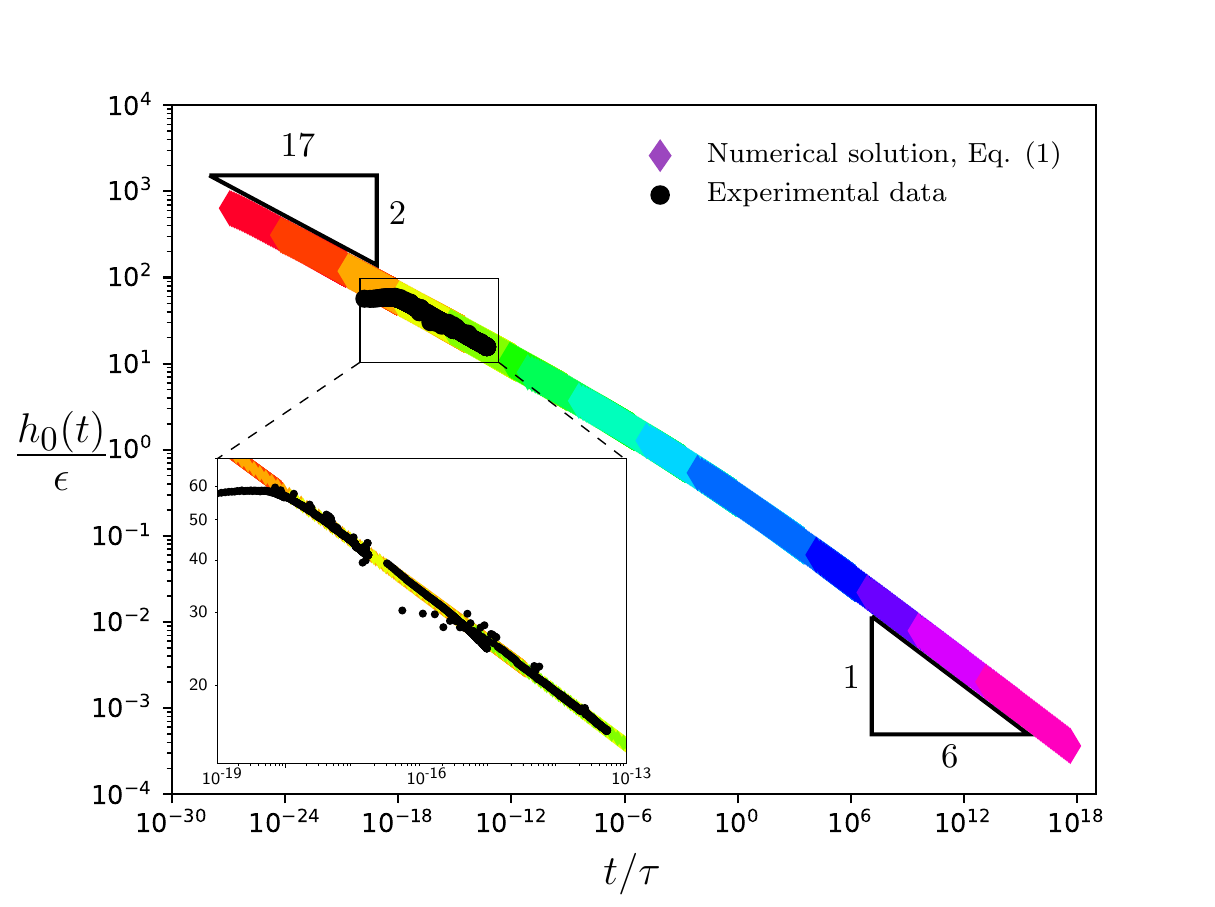}
\caption{Non-dimensional bump height as a function of dimensionless time in the bending case, for various initial values of $h_{\textrm{i}}/\epsilon\in[10^{-2};10^3]$. The coloured diamond-shaped markers are rescaled data points from the numerical solutions of the dimensionless version of Eq.~\eqref{eq:evolution_equation} with $N=0$, and the black circle-shaped markers are scaled experimental data points. The exponents of the two asymptotic regimes of Eqs.~\eqref{eq:relaxation}~and~\eqref{eq:relaxation2} are indicated with triangles.
The inset provides a zoom in the region containing the experimental data for the three samples, with initial aspect ratios $h_{\textrm{i}}/\epsilon=30$, 43, 56; corresponding respectively to $\epsilon=50$, 67, 26~nm; $h_{\textrm{i}}=1.52$, 2.9, 1.48~$\mu$m; and $R_{\textrm{i}}=9.6$, 16.5, 9.9 $\mu$m. The uncertainties in all experimental length scales are about 5\%. To compensate for those, the characteristic time $\tau$ for each sample is multiplied by a free fitting factor $\alpha=0.7$, 0.13, and 1.3, respectively.}
\label{fig:regime_transition}
\end{figure}

We first start by investigating the elastohydrodynamic leveling in the absence of thermal fluctuations ($N=0$). In Fig.~\ref{fig:num_vs_exp} we show the numerical solutions of the dimensionless version of Eq.~\eqref{eq:evolution_equation} for $N=0$ and we can see that the aspect ratio $h_{\textrm{i}}/\epsilon$ controls both the time scale for leveling and the detailed features of height profile. The smaller $h_{\textrm{i}}/\epsilon$, the faster the dimensionless leveling process. Also, the dip created near the advancing front of the perturbation is enhanced both in magnitude and lateral extent for smaller $h_{\textrm{i}}/\epsilon$. We remark that for each initial aspect ratio there is a transition period of a few numerical time steps preceding the onset of the leveling process. This part of the data is not included in Fig.~\ref{fig:regime_transition} as it is considered to depend on the initial condition, but does not influence the later dynamics.  For $h_{\textrm{i}}/\epsilon=43$, the numerical height profiles are further compared with our experiments, which are found to be in good agreement. We recall here that the elastic plate is floating on the liquid film and has edges that are free to move. Therefore, the pressure contribution from bending still largely dominates any contribution from stretching and Eq.~\eqref{eq:evolution_equation} is still valid.

We now turn to a scaling analysis of Eq.~\eqref{eq:evolution_equation} for $N=0$. When $h_0(t)/\epsilon\ll 1$, the equation can be linearized and reduces to $12\mu\partial h/\partial t=B\epsilon^3\partial^{6}h/\partial x^6$ and we deduce the long-term scaling for the temporal evolution of the horizontal length of the bump: $R(t)\sim [B\epsilon^3t/(12\mu)]^{1/6}$. Since there is area conservation in the $(x,z)$-plane, we assume $R(t)h_0(t)$ to be constant, that is evaluated to $R_{\textrm{i}}h_{\textrm{i}}$ at $t=0$. By combining these scaling relations we get for $h_0(t)/\epsilon\ll 1$
\begin{equation}
\frac{h_0(t)}{\epsilon}\sim\left(\frac{\tau}{t}\right)^{1/6}\ ,
\label{eq:relaxation}
\end{equation}
where $\tau = 12\mu h_{\textrm{i}}^{\,6}R_{\textrm{i}}^{\,6}/(B \epsilon^9)$ is the characteristic time scale for the bending-driven leveling dynamics. As we operate within the regime where bending dominates over stretching, a similar result is obtained by considering the force balance between the viscous and bending forces~\cite{Vandeparre2010}. Also if we include isotropic stretching due to clamped boundaries a similar scaling law appear, but now with an additional logarithmic term, $R(t)\sim (t/\textrm{log}(t))^{1/6}$~\cite{Kodio2017}. However, when $h_0(t)/\epsilon\gg 1$ we must match the curvature of a traveling-wave solution localized near the advancing front with the quasi-static solution to obtain the correct scaling \cite{lister2013viscous}, \textit{i.e.}, constant pressure in the bump, leading to~\cite{carlson2018fluctuation}
\begin{equation}
\frac{h_0(t)}{\epsilon}\sim\left(\frac{\tau}{t}\right)^{2/17}\ .
\label{eq:relaxation2}
\end{equation}
By balancing the two asymptotic predictions above, we expect the crossover between them to occur around $t/\tau\approx1$. In addition, these asymptotic regimes suggest that $h_0(t)/\epsilon$ is essentially a function of $t/\tau$ only, independent of the value of $h_{\textrm{i}}/\epsilon$ in particular.

In order to test our scaling predictions, we compute numerical solutions of the dimensionless version of Eq.~\eqref{eq:evolution_equation} for $N=0$, with $h_{\textrm{i}}/\epsilon\in[10^{-2}, 10^{3}]$ and extract $h_0(t)/\epsilon$ as a function of $t/\tau$. These numerical results are plotted in Fig.~\ref{fig:regime_transition} and compared to the experimental data. For each sample, the experimental data is matched to the numerical data through one fitting parameter $\alpha$ in front of the time scale $\tau$. The values of the fluid viscosity and elastic Young modulus are highly sensitive to the temperature in the experiments, and we estimate them to be $\mu\approx10^4\;\textrm{Pa}\;\textrm{s}$~\cite{mcgraw2012, ilton2016} and $E\approx2.6$~GPa~\cite{springer}, respectively. Since all experiments were carried out at the same temperature and with the same polymer, sample-to-sample variations in $\tau$ result only from uncertainties in the geometrical parameters $h_{\textrm{i}}$, $R_{\textrm{i}}$, $d$, and $\epsilon$. The obtained $\alpha$ values are 0.13, 0.7, and 1.3 for the three samples and each of these values are reasonably close to unity. More importantly, the sample-to-sample variations in $\alpha$ do not exceed a factor of 10, which is well within the expected relative error arising from the high sensitivity of $\tau$ to the geometrical parameters. The general agreement between the experimental data and the numerical predictions is good, over about 5 orders of magnitude in $t/\tau$. The systematic early-time tail in the experimental data might be attributed to the initial compressive thermal stresses in the elastic layer, which arise due to the rapid heating of the samples from room temperature to $T = 130^{\circ}$C, which relax prior to leveling and the time needed for the initial shape to enter the asymptotic regime.

The master curve in Fig.~\ref{fig:regime_transition} confirms that $h_0(t)/\epsilon$ is a function of $t/\tau$. Furthermore, the two scaling regimes predicted above are indeed present, with pre-factors close to unity, and the crossover between the two being located near $t/\tau\approx1$ as predicted. Any bump that initially starts in a thin pre-wetted film regime $h_0(t)/\epsilon\gg1$ will eventually crossover to a thick-film regime $h_0(t)/\epsilon\ll 1$, with the corresponding power laws in time. As a final remark, a similar combination (not included here) of numerical simulations and scaling analysis can been performed for an axisymmetric geometry, leading to $h_0(t)\sim t^{-2/11}$ for $h_{\textrm{i}}/\epsilon \gg 1$, and $h_0(t)\sim t^{-1/3}$ for $h_{\textrm{i}}/\epsilon \ll 1$.

\subsection{Stochastic leveling}
\begin{figure}
\includegraphics[trim= 10 3 10 10, clip, width=1\linewidth]{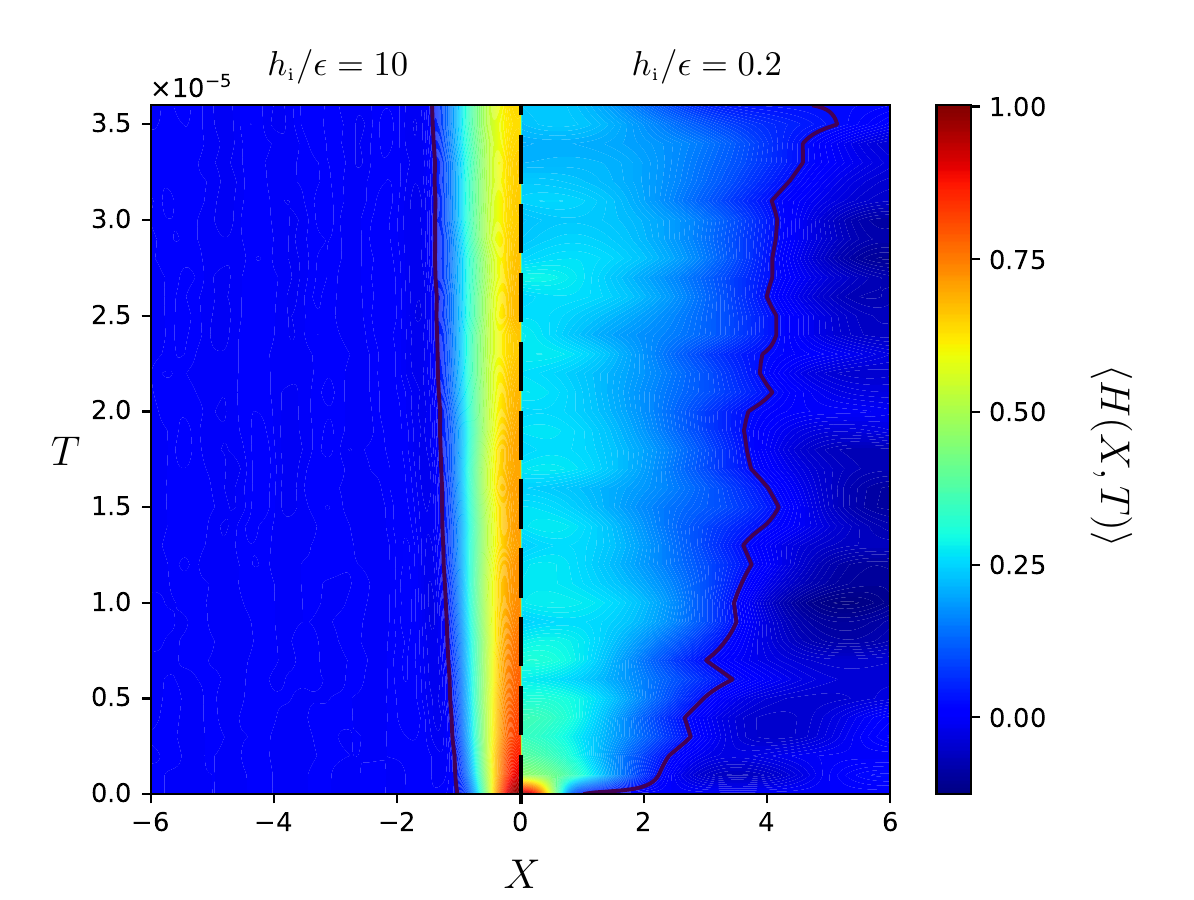}
\caption{Contour plot of the dimensionless bump height profile $\langle H(X,T) \rangle$ as a function of both the dimensionless position $X$ and time $T$, as obtained from numerical solutions of the dimensionless version (see text) of Eq.~\eqref{eq:evolution_equation}, with $N=5$, and for $h_{\textrm{i}}/\epsilon=10$ (left) or $h_{\textrm{i}}/\epsilon= 0.2$ (right). The thick solid lines indicate $\langle H(X,T)\rangle=0.03$ as an arbitrary reference.}
\label{fig:thermoplot}
\end{figure}

Next we investigate the leveling process when it is dominated by thermal fluctuations ($N> 0$). As shown in Fig.~\ref{fig:thermoplot}, the numerical solutions suggest that the aspect ratio $h_{\textrm{i}}/\epsilon$ is again essential, as it sets the time scale for leveling where the smaller $h_{\textrm{i}}/\epsilon$, the faster the dimensionless leveling process. Moreover, by comparison with the deterministic ($N=0$) case in Fig.~\ref{fig:num_vs_exp}, the stochastic ($N> 0$) profiles exhibit spatiotemporal fluctuations and adopt different average shapes and leveling dynamics.
\begin{figure}
\includegraphics[trim= 3 3 27 20, clip, width=1\linewidth]{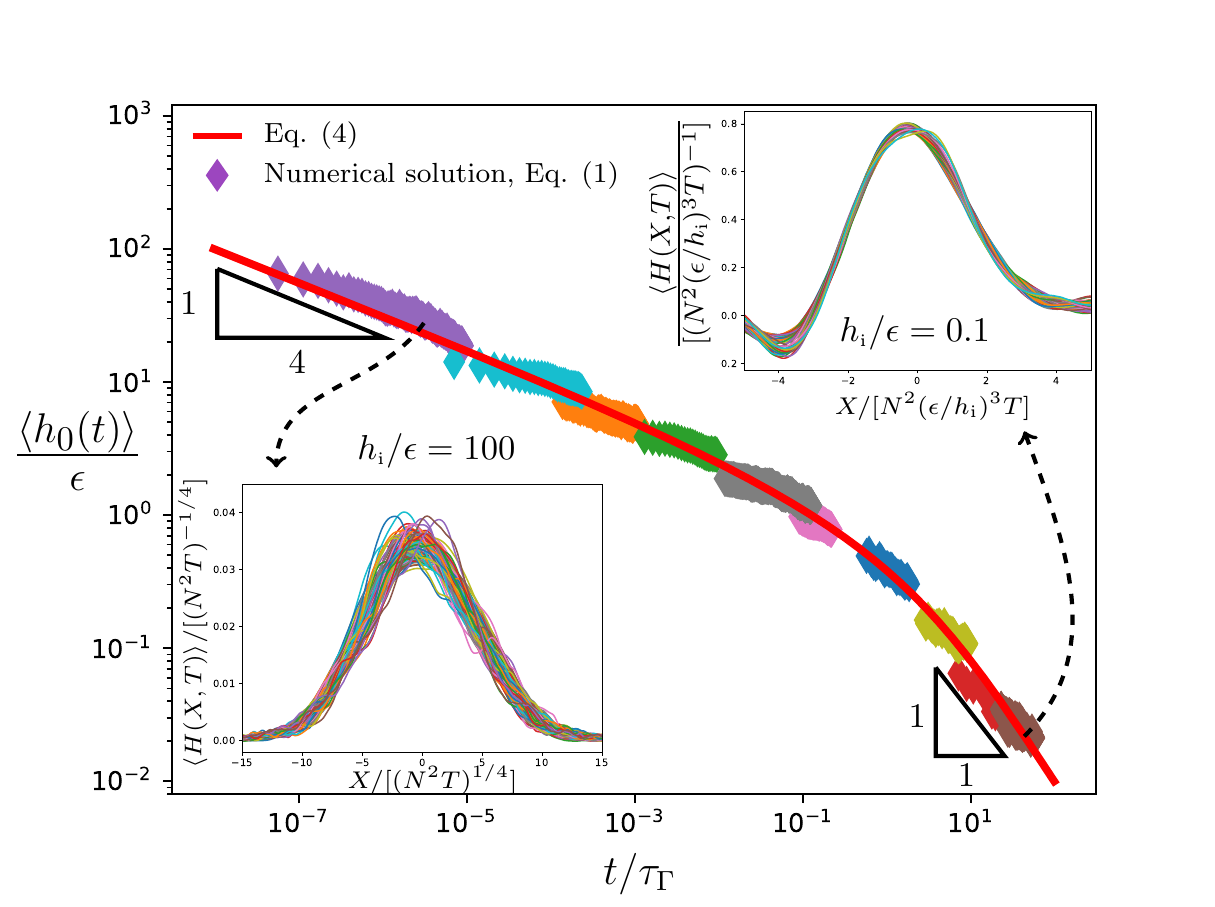}
\caption{Non-dimensional bump height as a function of dimensionless time in the stochastic leveling dynamics, for different initial values of $h_{\textrm{i}}/\epsilon\in[10^{-1}, 10^2]$. The coloured diamond-shaped markers are rescaled data points from the numerical solutions of Eq.~\eqref{eq:evolution_equation} with $5\leq N\leq 8$. Each data set is an average from minimum 30 numerical solutions. The solid red line corresponds to Eq.~\eqref{eq:evolution_thermal} with a prefactor of order unity. The insets show rescaled bump height profiles for $h_{\textrm{i}}/\epsilon=100$ with $T\in[1,10]\times 10^{-3}$ (lower left); and for $h_{\textrm{i}}/\epsilon=0.1$ with $T\in[4,4.6]\times 10^{-3}$(upper right).}
\label{fig:masterfluc}
\end{figure}

To go further, we propose a scaling analysis of Eq.~\eqref{eq:evolution_equation}, inspired by Ref.~\cite{cormier2012beyond}. We consider specifically the $N\gg1$ limit, for which the thermal fluctuations are the dominant driving contribution to the dynamics and we assume that we can neglect the bending term so that Eq.~\eqref{eq:evolution_equation} reduces to $\partial h/\partial t= \Gamma \partial[(\epsilon+h)^{3/2}\eta]/\partial x$. We consider the average quantities $\langle h_0(t)\rangle $ and $\langle R(t)\rangle $, where we invoke the $\sim (tx)^{-1/2}$ scaling~\cite{davidovitch2005spreading} for the root mean square value of the averaged noise over a space interval $x$ and a time interval $t$. By assuming that the average area conservation in the $(x,z)$-plan can be expressed as $\langle h_0(t)\rangle \langle R(t)\rangle \sim h_{\textrm{i}}R_{\textrm{i}}$, we get
\begin{equation}
\frac{\langle h_0(t)\rangle}{\epsilon}\left[1+\frac{\langle h_0(t)\rangle}{\epsilon}\right]^3\sim\frac{\tau_{\Gamma}}{t}\
\label{eq:evolution_thermal}
\end{equation}
where $\tau_{\Gamma} = 6\mu h_{\textrm{i}}^{\,3}R_{\textrm{i}}^{\,3} b/(k_{\textrm{B}} T_{\textrm{A}}\epsilon^4)$ is the characteristic time scale for the stochastic leveling dynamics. Interestingly, Eq.~\eqref{eq:evolution_thermal} describes a complete crossover between two asymptotic regimes in the stochastic leveling dynamics: for $\langle h_0(t)\rangle/\epsilon\gg 1$, we obtain $\langle h_0(t)\rangle/\epsilon\sim(\tau_{\Gamma}/t)^{1/4}$, and thus we recover $\langle h_0(t)\rangle\sim t^{-1/4}$~\cite{davidovitch2005spreading}; while for $\langle h_0(t)\rangle/\epsilon\ll 1$, we get $\langle h_0(t)\rangle/\epsilon\sim\tau_{\Gamma}/t$. We expect the crossover between the two asymptotic regimes to occur around $\langle h_0(t)\rangle/\epsilon\approx 1$, \textit{i.e.}, around $t/\tau\approx1/8$.

In order to test the prediction in Eq. (4), we compute the numerical solution of the dimensionless version of Eq.~\eqref{eq:evolution_equation} for $5\leq N\leq 8$, with $h_{\textrm{i}}/\epsilon\in[10^{-1}, 10^{2}]$. By averaging over a minimum of 30 realizations, we can extract $\langle h_0(t)\rangle/\epsilon$ as a function of $t/\tau_{\Gamma}$ and the results are plotted in Fig.~\ref{fig:masterfluc}. The data from the numerical solutions is in good agreement with Eq.~\eqref{eq:evolution_thermal} for all $\langle h_0(t)\rangle/\epsilon$, with no adjustable parameter. Our results highlight that Eq.~\eqref{eq:evolution_thermal} gives an accurate prediction of the stochastic leveling dynamics and show that the missing pre-factor is close to unity. Finally, in order to further highlight the underlying self-similarity associated with each of the two asymptotic regimes, the insets of Fig.~\ref{fig:masterfluc} show the corresponding bump height profiles rescaled according to  Eq.~\eqref{eq:evolution_thermal}. In each asymptotic regime the height profiles collapse onto a universal shape which confirms the overall self-similarity in the leveling dynamics.

\section{CONCLUSION}
We have described the elastohydrodynamic and stochastic leveling of an elastic plate placed atop a viscous film. By combining numerical solutions, scaling analysis, and experiments, we identified various canonical regimes. Our results highlight the importance of the driving mechanism, either by elastic bending of the plate or thermal fluctuations, and the influence of the aspect ratio of the bump height to the film height. For each of these two driving mechanisms, a crossover between two distinct asymptotic regimes is controlled by the aspect ratio. These findings can be helpful to explain elastohydrodynamic leveling dynamics found in biological, engineering, or geological processes.

\begin{acknowledgments}
The financial support by the Research Council of Norway (project grant
263056), the Natural Science and Engineering Research Council of Canada and the Joliot Chair of ESPCI Paris is gratefully acknowledged. The authors also thank Tak Shing Chan, Miroslav Kuchta, Corentin Mailliet, Ren\'e Ledesma-Alonso, Maxence Arutkin, Elie Rapha\"el, Howard Stone, Pauline Olive, Katelyn Dixon, Paul Fowler, Mark Ilton and David Dean, for interesting discussions and preliminary works in different geometries.
\end{acknowledgments}

\bibliographystyle{unsrt}
\bibliography{Pedersen2019}
\end{document}